# Tunneling magnetoresistance in magnetic tunnel junctions with a single ferromagnetic electrode

Kartik Samanta,[1,*,†] Yuan-Yuan Jiang,[2,3,*] Tula R. Paudel,[4] Ding-Fu Shao,[2,‡] and Evgeny Y. Tsymbal[1,§]

[1] *Department of Physics and Astronomy & Nebraska Center for Materials and Nanoscience, University of Nebraska, Lincoln, Nebraska 68588, USA*

[2] *Key Laboratory of Materials Physics, Institute of Solid-State Physics, HFIPS, Chinese Academy of Sciences, Hefei 230031, China*

[3] *University of Science and Technology of China, Hefei 230026, China*

[4] *Department of Physics, South Dakota School of Mines and Technology, Rapid City, South Dakota 57701, USA*

Magnetic tunnel junctions (MTJs) are key components of spintronic devices, such as magnetic random-access memories. Normally, MTJs consist of two ferromagnetic (FM) electrodes separated by an insulating barrier layer. Their key functional property is tunneling magnetoresistance (TMR) that is a change in MTJ's resistance when magnetization of the two electrodes alters from parallel to antiparallel. Here, we demonstrate that TMR can occur in MTJs with a single FM electrode, provided that the counter electrode is an antiferromagnetic (AFM) metal that supports a spin-split band structure and/or a Néel spin current. Using $RuO_2$ as a representative example of such antiferromagnet and $CrO_2$ as a FM metal, we design all-rutile $RuO_2/TiO_2/CrO_2$ MTJs to reveal a non-vanishing TMR. Our first-principles calculations predict that magnetization reversal in $CrO_2$ significantly changes conductance of the MTJs stacked in the (110) or (001) planes. The predicted giant TMR effect of about 1000% in the (110) oriented MTJs stems from spin-dependent conduction channels in $CrO_2$ (110) and $RuO_2$ (110), whose matching alters with $CrO_2$ magnetization orientation, while TMR in the (001) oriented MTJs originates from the Néel spin currents and different effective $TiO_2$ barrier thickness for the two magnetic sublattices that can be engineered by the alternating deposition of $TiO_2$ and $CrO_2$ monolayers. Our results demonstrate a possibility of a sizable TMR in MTJs with a single FM electrode and offer a practical test for using the altermagnet $RuO_2$ in functional spintronic devices.

Spintronics utilizes a spin degree of freedom in electronic devices to encode information [1]. A typical and widely used spintronic device is a magnetic tunnel junction (MTJ), which is composed of two ferromagnetic (FM) metal electrodes separated by a non-magnetic insulating tunnel barrier [2-6]. The key functional property of an MTJ is tunneling magnetoresistance (TMR) that is a change of MTJ's resistance in response to magnetization reversal of the two FM electrodes from parallel to antiparallel [7]. The TMR effect can be as high as a few hundred percent [5,6] allowing the use of MTJs as building blocks of magnetic random-access memories (MRAMs) [8].

The physics of TMR has been well understood in terms of spin-dependent tunneling that is controlled by the spin-polarized electronic band structure of ferromagnets and evanescent states of the tunneling barrier. In a crystalline MTJ, where the transverse wave vector is conserved in the tunneling process, wave functions belong to the symmetry group of the wave vector of the whole MTJ. This entails symmetry matching of the incoming and outcoming Bloch states in the electrodes and evanescent states in the barrier [9]. In particular, matching of the majority-spin $\Delta_1$ band in the Fe (001) electrode to the $\Delta_1$ evanescent state in the MgO (001) barrier layer is responsible for a large positive spin polarization and giant values of TMR predicted for crystalline Fe/MgO/Fe (001) MTJs [10]. Also, the complex band structure explains a large *negative* spin polarization of electrons tunneling from FM bcc Co (001) through an $SrTiO_3$ (001) tunneling barrier [11] consistent with the experimental observations [12,13]. It is now commonly accepted that the transport spin polarization of MTJs is controlled by the ferromagnet/barrier pair rather than the ferromagnet alone, which can be understood in terms of the interface transmission function [14].

In a two-terminal device, such as an MTJ, the spin polarization of the tunneling current cannot be detected on its own but requires a *magnetic* counter electrode to measure TMR. This is because in a tunnel junction with a *non-magnetic* counter electrode, time reversal operation $T$ flips the magnetization of the FM electrode and reverses the current direction but does not change the conductance magnitude, even in the presence of spin-orbit coupling [15]. While ferromagnets are commonly used as counter electrodes in MTJs, the question arises if an antiferromagnet could be used instead to detect the tunneling spin polarization generated by a ferromagnet/barrier pair. This question is interesting not only from the fundamental point of view but also from the practical perspective, since in conventional MTJs, magnetization pinning of the counter FM electrode (a pinned layer) is often required, which is typically achieved using an exchange bias provided by an additional antiferromagnet (a pinning layer). Using an antiferromagnetic (AFM) counter electrode instead would not require a pinning layer, simplifying the MTJ structure.

In this paper, we propose two strategies to realize an MTJ with a single FM electrode. The first approach exploits a low-symmetry oriented AFM counter electrode that exhibits a spin-split band structure and uncompensated momentum-dependent transport spin polarization [16]. The second approach employs an AFM metal with a strong intra-sublattice coupling, revealing the staggered Néel spin current. To demonstrate these strategies, we consider $RuO_2$, a high-Néel-temperature AFM metal, as a counter electrode in all-rutile MTJs with a $CrO_2$ FM electrode and $TiO_2$ tunneling barrier. This choice of an AFM electrode is driven by the fact that $RuO_2$ supports a spin polarized current in the [110] direction and a staggered Néel spin current in the [001] direction [17]. Using first-principles quantum-transport calculations, we predict sizable TMR for $RuO_2/TiO_2/CrO_2$ (110) and (001) MTJs.

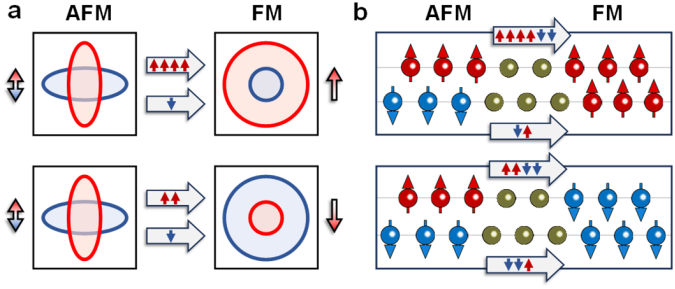

FIG. 1: Schematics of TMR in MTJs with a single FM electrode and an AFM counter electrode. (**a**) TMR due to the anisotropic Fermi surface along a low-symmetry transport direction in an AFM electrode that makes the electric current spin-polarized. This represents the [110] direction in $RuO_2$. Depending on the magnetization orientation of the FM layer (indicated by arrows) with respect to the Néel vector of the AFM layer (indicated by double arrows), transmission is high (top panel) or low (bottom panel) resulting in TMR. (**b**) TMR due to the Néel spin current along a high-symmetry direction of the AFM electrode. This represents the [001] direction in $RuO_2$. Different transmission of an MTJ for FM magnetization pointing up (top panel) and down (bottom panel), resulting in TMR, originates from different effective barrier thickness between two magnetic sublattices in the AFM electrode and the FM metal lattice.

In crystalline MTJs, TMR is determined by the momentum-dependent transport spin polarization $p_\parallel(\vec{k}_\parallel)$ of the two electrodes, where $\vec{k}_\parallel$ is the wave vector that is transverse to the transport direction. An FM electrode hosts unbalanced $p_\parallel(\vec{k}_\parallel)$ resulting in a finite net spin polarization. To employ an AFM metal as a counter electrode in an MTJ, this antiferromagnet should also have unbalanced $p_\parallel(\vec{k}_\parallel)$ along the transport direction; otherwise, magnetization reversal would just flip spin contributions to MTJ's conductance without changing their magnitudes. Most compensated antiferromagnets, however, exhibit $PT$ or $U\tau$ symmetries (where $P$ is space inversion, $U$ is spin flip, and $\tau$ is half a unit cell translation) that not only prevent net magnetization but also enforce a spin-degenerate band structure and hence vanishing $p_\parallel(\vec{k}_\parallel)$. Thus, the desired AFM electrode must belong to a magnetic space group which does not have $PT$ and $U\tau$ among their symmetry operations. Among such antiferromagnets are certain types of collinear antiferromagnets [18 - 23], dubbed altermagnets [24, 25], and noncollinear antiferromagnets [26,27]. These non-relativistically spin-split antiferromagnets have been proposed for and utilized in AFM tunnel junctions (AFMTJs) [22, 26-34]. Such AFM metals allow for non-zero net spin polarization like ferromagnets. This behavior is illustrated in Fig. 1(a), showing a spin-dependent Fermi surface of an antiferromagnet providing an unbalanced $p_\parallel(\vec{k}_\parallel)$ along the transport direction and hence a globally spin-polarized current resulting in a non-zero TMR in an MTJ with a single FM electrode.

Another strategy is to use a spin-degenerate antiferromagnet or a spin-split antiferromagnet with high-symmetry layer stacking that supports a Néel spin current (i.e., a staggered spin current on the two magnetic sublattices) [17]. While purely bulk-based considerations do not support TMR in an MTJ with a single FM electrode in the case of these types of AFM counter electrodes, certain kinds of *engineered* high-quality epitaxial MTJs may provide conditions for a non-zero TMR. Specifically, in MTJs where epitaxial layer-by-layer growth occurs through an alternating addition of atoms to an atomic chain connecting each magnetic sublattice in the AFM electrode to the FM electrode lattice, the effective barrier thickness for the two magnetic sublattices can be engineered to be unequal. This behavior is schematically depicted in Figure 1(b), where two magnetic sublattices in the AFM electrode carry Néel spin currents which further propagate across the barrier into an FM metal. Due to the different barrier thickness for the two magnetic sublattices and the electric currents flowing in parallel, the resulting TMR is non-zero.

The recently discovered AFM metal $RuO_2$ (Fig. 1 (a)) [35, 36] supports a spin-polarized current along the [110] direction [22] and a Néel spin current along the [001] direction [17], and hence can serve as a counter electrode in an MTJ with a single FM electrode. $RuO_2$ has a rutile structure with two AFM sublattices $Ru_A$ and $Ru_B$ (Fig. 1(a)). The Néel vector is pointing along the [001] direction, and the Néel temperature is reported to be above 300 K [35]. The required properties of $RuO_2$ originate from its magnetic space group $P4_2'/mnm'$ that has broken $PT$ and $U\tau$ symmetries, supporting spin splitting of the band structure in the absence of spin-orbit coupling. As is evident from Figure 2 (c), the energy bands of bulk $RuO_2$ have a pronounced spin splitting along the high-symmtry Γ-M and Z-A lines, whereas they are spin-degenerate along the Γ-X, Γ-Z, X-M, Z-R, and R-A lines [37]. This fact indicates spin-polarized transport along the [110] direction and non-spin-polarized transport along the [001] direction in $RuO_2$. There is, however, a strong intra sub-lattice coupling along the [001] direction which supports a Néel spin current that can be used for TMR [17].

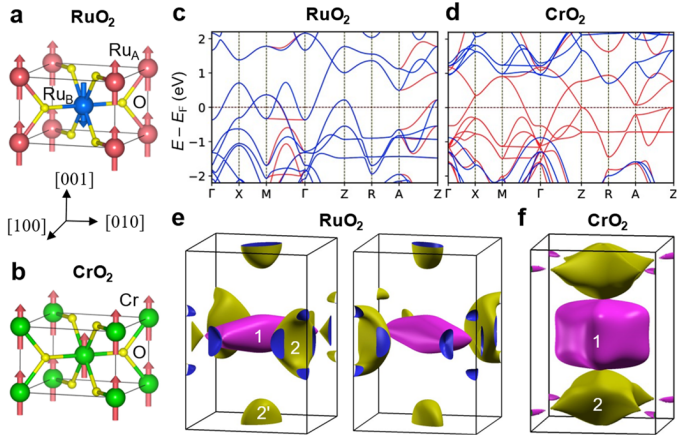

FIG. 2: (**a**, **b**) Atomic and magnetic structure of $RuO_2$ (a) and $CrO_2$ (b). (**c**, **d**) Band structure of $RuO_2$ (c) and $CrO_2$ (d). Red and blue curves indicate up- and down-spin bands, respectively. (**e**, **f**) Fermi surfaces of $RuO_2$ (e) and $CrO_2$ with essential bands numbered (f).

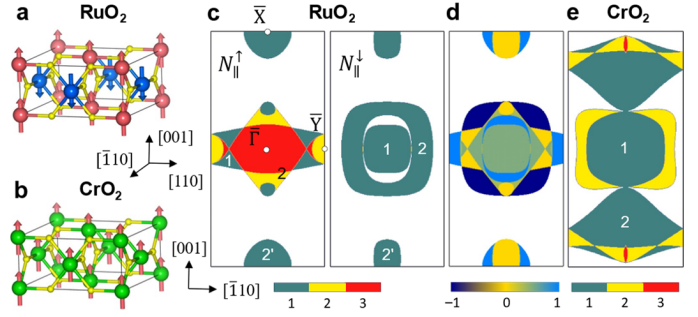

FIG. 3: (**a**, **b**) Supercells of $RuO_2$ (110) (a) and $CrO_2$ (110) (b). (**c**) The distribution of conduction channels in the 2DBZ for up-spin (left) and down-spin (right) of $RuO_2$ (110). High symmetry points in the 2DBZ are indicated, assuming that $z \parallel [110]$, and essential contributing bands numbered. (**d**) Spin polarization of conduction channels in $RuO_2$ (110). White color indicates absent conduction channels. (**e**) Same as (c) for up-spin of $CrO_2$ (110).

As an FM electrode, we consider $CrO_2$ that has a rutile structure (Fig. 1(b)) at ambient conditions and belongs to space group $P4_2/mnm$ [38]. $CrO_2$ is an FM metal with the Curie temperature of 385-400K [39]. As is evident from Fig. 1(d), the majority-spin bands of bulk $CrO_2$ cross the Fermi-energy while the minority-spin bands have a band gap that signifies half-metallicity of $CrO_2$ [40, 41] and the associated integer magnetization of 2 $\mu_B$/f.u. in the ground state. We note here that the half-metallic nature of $CrO_2$ is not essential for the TMR effects predicted in this paper.

The rutile space group $P4_2/mnm$ has four-fold rotational symmetry $C_4$ with respect to the [001] axis. While this symmetry is preserved by magnetism in $CrO_2$, it is broken in $RuO_2$. This is reflected in the Fermi surfaces of bulk $RuO_2$ and $CrO_2$. The Fermi surface of $RuO_2$ is spin-split such that the up- and down-spin Fermi surfaces can be transformed to each other by a 90° rotation around the [001] axis (Fig. 2(e)). In contrast, the up-spin Fermi surface of $CrO_2$ has four-fold rotational symmetry with respect to the [001] axis (Fig. 2(f)). Note that the down-spin Fermi-surface does not exist due to half-metallicity of $CrO_2$. As a result of these bulk symmetries, combining $RuO_2$ (001) and $CrO_2$ (001) as electrodes in an MTJ stacked in the (001) plane is not expected to produce TMR. On the contrary, transport along the {110} direction is expected to be spin polarized resulting in a non-zero TMR effect.

The latter fact is evident from the calculated number of conduction channels of bulk $RuO_2$ and $CrO_2$ along the transport direction, *i.e.*, the number of propagating Bloch states in the momentum space [37]. For the [001] transport direction in $RuO_2$, the distribution of conduction channels for up-spin ($N_\parallel^\uparrow$) and down-spin ($N_\parallel^\downarrow$) electrons as a function of transverse wave vector $\vec{k}_\parallel$ has congruent shapes in the two-dimensional Brillouin zone (2DBZ). $N_\parallel^\uparrow$ and $N_\parallel^\downarrow$ can be transformed to each other by a 90° rotation around the $\bar{\Gamma}$ point [37], reflecting the respective property of the $RuO_2$ Fermi surface (Fig. 2(e)). At the same time, the distribution of conduction channels in $CrO_2$ (001) has a four-fold rotational symmetry inherited from its Fermi surface (Fig. 2(f)). As a result, an MTJ combining AFM $RuO_2$ and FM $CrO_2$ electrodes is non-expected to produce TMR along the [001] transport direction, since total transmission of the MTJ with opposite magnetization directions are to be the same.

In contrast, due to the spin splitting along the $\Gamma$-M line (Fig. 2(c)), $RuO_2$ is spin-polarized along the [110] direction. This is seen from the calculated distribution of conduction channels, $N_\parallel^\uparrow$ and $N_\parallel^\downarrow$, shown in Fig. 3(c), where a $RuO_2$ (110) supercell is used in the calculation (Fig. 3(a)). It is evident that for up-spin electrons (Fig. 3(a), left), there is an elliptic electron pocket elongated in the $\bar{\Gamma}$ - $\bar{Y}$ direction (band 1 in Fig. 3(c)) and overlapped with a rhombic hole pocket around the $\bar{\Gamma}$ point (band 2), and a small pocket at the $\bar{Y}$ point. Band 2 has two conduction channels on its own, which results in $N_\parallel^\uparrow = 3$ in the regions of overlap with band 1. There is also a small hole pocket of $N_\parallel^\uparrow = 1$ at the $\bar{X}$ point (band 2′). The same kind of Fermi surface sheets, but rotated by 90° around the [001] axis, contribute to down-spin conduction channels of $RuO_2$ (Fig. 3(a), right). Due to no overlap between their projections, $N_\parallel^\downarrow = 1$ in all regions of the 2DBZ where these bands appear. For $CrO_2$ (110), only up-spin Bloch states are present. As seen from Fig. 3(e), there is a large electron pocket around the 2DBZ center (band 1) with $N_\parallel^\uparrow = 1$ that alters to $N_\parallel^\uparrow = 2$ closer to the $\bar{Y}$ point. There is also a large hole pocket at the $\bar{X}$ point (band 2). This distribution of conduction channels for $RuO_2$ (110) and $CrO_2$ (110) is consistent with the band-decomposed Fermi surfaces [37].

The unbalanced distribution of $N_\parallel^\uparrow$ and $N_\parallel^\downarrow$ in RuO$_2$ (110) leads to $\vec{k}_\parallel$-dependent spin polarization $p_\parallel(\vec{k}_\parallel) = \frac{N_\parallel^\uparrow - N_\parallel^\downarrow}{N_\parallel^\uparrow + N_\parallel^\downarrow}$ and non-zero net spin polarization $p = \frac{\sum N_\parallel^\uparrow - \sum N_\parallel^\downarrow}{\sum N_\parallel^\uparrow + \sum N_\parallel^\downarrow}$. As seen from Fig. 3(c), there is a full spin polarization ($p_\parallel = \pm 100\%$) in the regions of a finite $N_\parallel^{\uparrow,\downarrow}$ for one spin channel and zero $N_\parallel^{\uparrow,\downarrow}$ for the other spin channel. Unlike RuO$_2$ (001), the total transport spin-polarization is non-vanishing for RuO$_2$ (110), namely $p = 31\%$, which is comparable to the spin polarization of representative FM metals like Fe, Co, and Ni [42, 43]. Thus, RuO$_2$ (110) can be used as a spin detector in MTJs with a single FM electrode.

To demonstrate this behavior, we construct an MTJ using an FM CrO$_2$ electrode, an AFM RuO$_2$ counter electrode, and a TiO$_2$ barrier layer. All constituents of this MTJ have the rutile structure and similar lattice constants [38, 44, 45], providing a possibility for epitaxial growth of the crystalline MTJ. We first consider a RuO$_2$/TiO$_2$/CrO$_2$ (110) MTJ where electrons are tunneling in the [110] direction. Figure 4(a) shows the atomic structure of the RuO$_2$/TiO$_2$/CrO$_2$ (110) supercell that is used in the transport calculations [37]. We find a wide band gap of TiO$_2$ which is well maintained in this MTJ with the Fermi energy $E_F$ located nearly in the middle (Fig. 4(b)). We define a parallel (P) state of the MTJ for Cr moments parallel to Ru$_A$ moments and an antiparallel (AP) state for antiparallel Cr and Ru$_A$ moments.

Figure 4 (c) shows the calculated $\vec{k}_\parallel$-resolved transmission for the P state of the MTJ, $T_P(\vec{k}_\parallel)$, and for the AP state, $T_{AP}(\vec{k}_\parallel)$. Due to CrO$_2$ being half metal, only up-spin electrons contribute to $T_P$ and down-spin electrons to $T_{AP}$. We find that $T_P(\vec{k}_\parallel)$ and $T_{AP}(\vec{k}_\parallel)$ mirror the distribution patterns of the RuO$_2$ (110) conduction channels, $N_\parallel^\uparrow$ and $N_\parallel^\downarrow$, respectively (compare Figs. 3(c) and 4(c)). For up-spin electrons, the largest contribution to $T_P(\vec{k}_\parallel)$ comes from band 1 at the Fermi surface, whereas other bands contribute modestly. In contrast, for down-spin electrons, band 1 is elongated in the transport direction and its contribution to the transmission is small. The largest contribution to $T_{AP}(\vec{k}_\parallel)$ comes from band 2 that has a rounded-square shape with a hole around the $\bar{\Gamma}$ point which is filled by another band.

It is notable that $T_P(\vec{k}_\parallel)$ has sizably reduced transmission at the $\bar{\Gamma}$ point and along the $\bar{\Gamma}$ - $\bar{X}$ line. This can be explained based on symmetry analysis. All tunneling states belong to the symmetry group of the wave vector of the whole MTJ. Along the [110] direction of the rutile structure, the symmetry group of the wave vector is equivalent to that of the $C_{2v}$ point group and has four irreducible representations: $\Sigma_1 (x^2, y^2, z^2)$, $\Sigma_2 (xy)$, $\Sigma_3 (xz)$, and $\Sigma_4 (yz)$, where in parentheses, we show basis functions generating these representations [46]. Up-spin bands 1 and 2 at the Fermi surface of RuO$_2$ are characterized, respectively, by the $\Sigma_3$ and $\Sigma_4$ symmetry representations, while

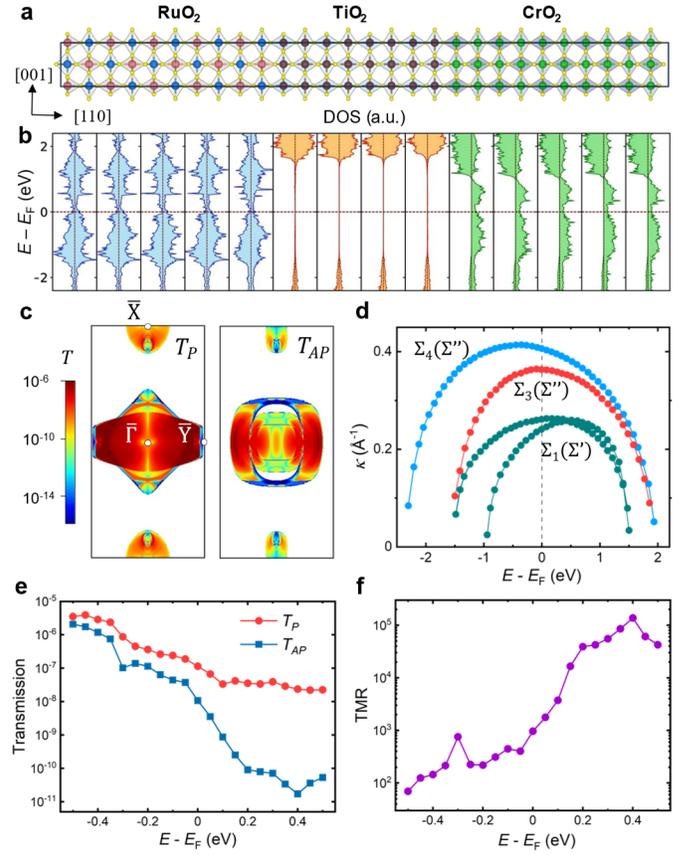

FIG. 4: (**a**) Atomic structure of RuO$_2$/TiO$_2$/CrO$_2$ (110) MTJ. (**b**) Calculated layer-resolved density of states (DOS) for the MTJ shown in (a). The horizontal line indicates the Fermi energy. (**c**) Calculated $\vec{k}_\parallel$-resolved transmission in the 2DBZ for P (left) and AP (right) states of the MTJ. (**d**) Decay rate $\kappa$ as a function of energy for four evanescent states in bulk TiO$_2$ (110) with the lowest $\kappa$. The vertical dashed line indicates the Fermi energy that is placed at the same position with respect to the band gap of TiO$_2$ as in the DOS (b). Band symmetries are displayed. (**e, f**) Calculated total transmissions, $T_P$ and $T_{AP}$, for P and AP states of the MTJ and TMR (c) as functions of energy.

CrO$_2$ band 1 has the $\Sigma_4$ character. As a result, at the $\bar{\Gamma}$ point of the 2DBZ, transmission from RuO$_2$ band 1 is forbidden by the symmetry mismatch of the incoming and outcoming Bloch states in the electrodes, while transmission from RuO$_2$ band 2 is allowed. However, this transmission is significantly suppressed by the evanescent states of TiO$_2$. As seen from Figure 4(d), the two evanescent states in TiO$_2$ (110) with the lowest decay rate $\kappa$ have the $\Sigma_1$ character and therefore do not support transport of electrons originating from the $\Sigma_4$ bands. The evanescent state of the $\Sigma_4$ symmetry has a much larger decay rate (Fig. 4(d)) and hence transmission at the $\bar{\Gamma}$ point stays small. Moving along the $\bar{\Gamma}$ - $\bar{X}$ line reduces symmetry of the wave vector to that equivalent to the $C_s$ point group (a subgroup of $C_{2v}$), where only two irreducible representations remain: $\Sigma' (x^2, y^2, z^2, xz)$ and $\Sigma''(xy, yz)$. As a result, the symmetry mismatch between the $\Sigma'$-symmetry band 1 in $RuO_2$ and $\Sigma''$-symmetry band 1 in $CrO_2$ remains as at the $\bar{\Gamma}$ point, while the transport from the $\Sigma''$-symmetry band 2 in $RuO_2$ should occur only through higher-$\kappa$ evanescent states because the lowest-$\kappa$ evanescent states in $TiO_2$ (110) have the $\Sigma''$ character.

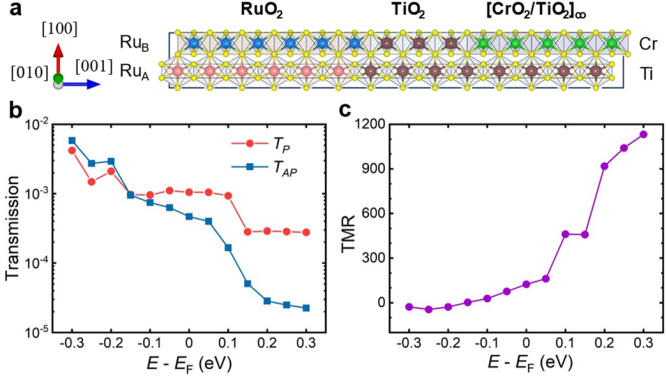

FIG. 5: (a) Atomic structure of $RuO_2/TiO_2/[TiO_2/CrO_2]_\infty$ (001) MTJ. (b,c) Calculated transmission, $T_P$ and $T_{AP}$, for magnetic moments of Cr atoms parallel and antiparallel to $Ru_A$ atoms (b) and TMR (c) as functions of energy for the MTJ shown in (a).

Moving away from the $\bar{\Gamma}$ - $\bar{X}$ line in the 2DBZ significantly enhances transmission (Fig. 4(c)). There are no symmetry restrictions, and transmission can occur through low-$\kappa$ states in $TiO_2$ involving $RuO_2$ band 1. The high transmission is supported by the distribution of the decay rate $\kappa(\vec{k}_\parallel)$ of the two lowest-$\kappa$ evanescent states in the 2DBZ [37]. The increase in transmission away from the $\bar{\Gamma}$ - $\bar{X}$ line is especially dramatic for the P-aligned MTJ (Fig. 4(c), left) due to a large overlap of the up-spin Fermi surface of $RuO_2$ with that of $CrO_2$ and three $RuO_2$ bands contributing to the transmission (Fig. 3(c), left).

By integrating over $\vec{k}_\parallel$, we find that total transmission $T_P$ is significantly greater than $T_{AP}$, leading to a giant TMR ratio $\frac{T_P - T_{AP}}{T_{AP}}$ of 965%. This value is comparable to the theoretically predicted [10] and larger that the measured [5, 6] values of TMR for well-known Fe/MgO/Fe (001) MTJs. Figure 4(c) shows total transmissions, $T_P$ and $T_{AP}$, as functions of energy $E$ for the $RuO_2/TiO_2/CrO_2$ (110) MTJ. It is seen that both $T_P$ and $T_{AP}$ decrease with increasing $E$, while $T_P$ being always greater than $T_{AP}$. This decrease originates from $\kappa(E)$ increasing with energy for the evanescent state with the lowest $\kappa$ near $E_F$ (Fig. 4(d)). $T_{AP}$ as a function of energy decreases notably faster than $T_P$ due to the reduced contribution from the $RuO_2$ hole pocket (band 2 in Fig. 3(c)) which shrinks at higher energies. This leads to the significant enhancement of TMR (Fig. 4(f)).

Contrary to $RuO_2$ (110), $RuO_2$ (001) supports only spin-neutral longitudinal currents. As a result (and as we have argued above), no TMR seems to appear in MTJs with $RuO_2$ (001) and FM electrodes, due to zero spin polarization of $RuO_2$ (001). However, rutile $MO_2$ ($M$ is a transition metal element) is composed of chains of edge-sharing $MO_6$ octahedra along the [001] direction, where the adjacent chains share common corners of the octahedra (Fig. 5(a)). This structural feature favors strong intra-chain transport, and hence staggered Néel spin currents in $RuO_2$ along the $Ru_A$ and $Ru_B$ chains [17]. Since such chains of octahedra are persistent across the interfaces in a perfectly epitaxial rutile heterostructure with the [001] growth direction (Fig. 5(a)), Néel spin currents are expected to dominate the spin-dependent transport properties of the rutile MTJ. This property allows engineering rutile MTJs that utilize $RuO_2$ (001) and FM electrodes and exhibit non-vanishing TMR.

Here, we consider a $RuO_2/TiO_2/[TiO_2/CrO_2]_n/CrO_2$ (001) MTJ, where $[TiO_2/CrO_2]_n$ represents a superlattice of alternating $TiO_2$ (001) and $CrO_2$ (001) monolayers with $n$ repeats. Such a superlattice can be fabricated using modern thin-film growth techniques [47,48]. The layer-by-layer growth of the superlattice provides alternating $TiO_6$ and $CrO_6$ octahedra chains. This leads to different effective barrier thickness for the Néel spin currents originating from $Ru_A$ and $Ru_B$ chains and generates TMR (Fig. 1(b)). For simplification, we assume a $RuO_2/TiO_2/[TiO_2/CrO_2]_\infty$ (001) MTJ, where the right electrode is an infinite $TiO_2/CrO_2$ superlattice ($n = \infty$) (Fig. 5(a)). We find a nearly half-metallic behavior of the superlattice [37], indicating that this MTJ can be considered as an extreme case of an MTJ with different effective barrier thickness for the two magnetic sublattices.

For such an MTJ with a 7-monolayer thick $TiO_2$ layer, $T_P$ appears to be more than a factor of two higher than $T_{AP}$, resulting in a sizable TMR ratio of 125%. Changing electron energy $E$ alters TMR, reflecting changes in the transport spin polarization of $RuO_2$ [37] and in the associated values of $T_P$ and $T_{AP}$ (Fig. 5(b)). As seen from Figure 5 (c), the TMR changes from small negative values at $E = E_F - 0.3$ eV to very large positive values exceeding 1000% at $E = E_F + 0.3$ eV, due to the enhancement of a Néel spin current by the increase of energy. We note that having $Ru_A$ atom at the left interface, while suppresses TMR due to different relative orientations of the two interfacial moments, does not influence our conclusions as the sizable TMR is largely controlled by the spin polarization of the Néel spin currents in $RuO_2$ [37]. We also note that the predicted TMR for the (001)-stacked rutile MTJ oscillates in its magnitude as a function of $TiO_2$ thickness in $RuO_2/TiO_2/CrO_2$ (001) MTJs [37], which can be verified experimentally provided layer-by-layer epitaxial growth of the MTJ.

Both approaches are feasible in practice. The first approach utilizing $RuO_2$ (110) in MTJs with a single FM electrode is more straightforward and can be employed in MTJs with barriers and FM electrodes different from $TiO_2$ and $CrO_2$. Compared to AFMTJs based on $RuO_2$ [17], it offers a simple practical test for using RuO$_2$ in functional spintronic devices due to simplicity of FM switching by an applied magnetic field. The second approach utilizing RuO$_2$ (001) requires a stringent control of the epitaxial layer-by-layer growth of the MTJ structure [47, 48]. Realizing this approach experimentally would provide direct evidence of the Néel spin currents. It also has an advantage of the perpendicular-to-plane magnetic anisotropy of RuO$_2$ (001) [35] desirable for high-density memory applications. In addition, this approach can be realized in 2D lateral MTJs with a bilayer A-type AFM electrode and a bilayer FM electrode, where the effective barrier width can be controlled independently for each layer by the recently developed edge-epitaxy technique [49-51]. We hope, therefore, that our theoretical predictions will stimulate experimental studies of the proposed MTJs and development of associated spintronic devices.

*Note added:* while finalizing our manuscript, we became aware of a relevant preprint posted recently on archive [52].

**ACKNOWLEDGMENTS**

This work was primarily supported by the Division of Materials Research of the National Science Foundation (NSF grant No. DMR-2316665). Research at HFIPS was supported by the National Natural Science Foundation of China (Grants Nos. 12274411, 12241405, and 52250418), and the CAS Project for Young Scientists in Basic Research No. YSBR-084. Computations were performed at the University of Nebraska-Lincoln Holand Computing Center and Hefei Advanced Computing Center.

\* These authors contributed equally to this work.

† ksamanta2@unl.edu
‡ dfshao@issp.ac.cn
§ tsymbal@unl.edu